\documentclass[runningheads]{llncs}

\usepackage{lipsum}
\usepackage{graphicx}
\usepackage{wrapfig}
\usepackage{color}
\usepackage[colorlinks=false]{hyperref}
\usepackage{soul}
\usepackage{amssymb}
\usepackage{mwe}
\usepackage{float}
\usepackage{array}
\usepackage[british]{babel}
\usepackage{multirow}
\usepackage{makecell}
\usepackage{caption}
\usepackage{subcaption}
\newcolumntype{L}[1]{>{\raggedright\let\newline\\\arraybackslash\hspace{0pt}}m{#1}}
\newcolumntype{C}[1]{>{\centering\let\newline\\\arraybackslash\hspace{0pt}}m{#1}}
\newcolumntype{R}[1]{>{\raggedleft\let\newline\\\arraybackslash\hspace{0pt}}m{#1}}

\begin{document}
\title{Entity Extraction from Wikipedia List Pages}
\author{
	Nicolas Heist\orcidID{0000-0002-4354-9138} \and
	Heiko Paulheim\orcidID{0000-0003-4386-8195}
}
\authorrunning{N. Heist, H. Paulheim}

\institute{
	Data and Web Science Group, University of Mannheim, Germany
	\email{\{nico,heiko\}@informatik.uni-mannheim.de}
}

\maketitle

\begin{abstract}
When it comes to factual knowledge about a wide range of domains, Wikipedia is often the prime source of information on the web. DBpedia and YAGO, as large cross-domain knowledge graphs, encode a subset of that knowledge by creating an entity for each page in Wikipedia, and connecting them through edges. It is well known, however, that Wikipedia-based knowledge graphs are far from complete. Especially, as Wikipedia's policies permit pages about subjects only if they have a certain popularity, such graphs tend to lack information about less well-known entities. Information about these entities is oftentimes available in the encyclopedia, but not represented as an individual page. In this paper, we present a two-phased approach for the extraction of entities from Wikipedia's list pages, which have proven to serve as a valuable source of information. In the first phase, we build a large taxonomy from categories and list pages with DBpedia as a backbone. With distant supervision, we extract training data for the identification of new entities in list pages that we use in the second phase to train a classification model. With this approach we extract over 700k new entities and extend DBpedia with 7.5M new type statements and 3.8M new facts of high precision.

\end{abstract}

\section{Introduction} \label{introduction}
Knowledge graphs like DBpedia \cite{lehmann2015dbpedia} and YAGO \cite{mahdisoltani2013yago3} contain huge amounts of high-quality data on various topical domains. Unfortunately, they are - as their application on real-world tasks show - far from complete: IBM's DeepQA system uses both of them to answer Jeopardy! questions \cite{kalyanpur2012structured}. While the component that uses this structured information gives correct answers 87\% of the time (compared to 70\% correctness of the complete system), it is only able to provide answers for 2.3\% of the questions posed to it. Given that they find in another analysis that around 96\% of the answers to a sample of 3,500 Jeopardy! questions can be answered with Wikipedia titles \cite{chu2012textual}, it is safe to say that there is a lot of information in Wikipedia yet to be extracted.

\begin{figure}[t]
    \centering
    \includegraphics[width=\textwidth]{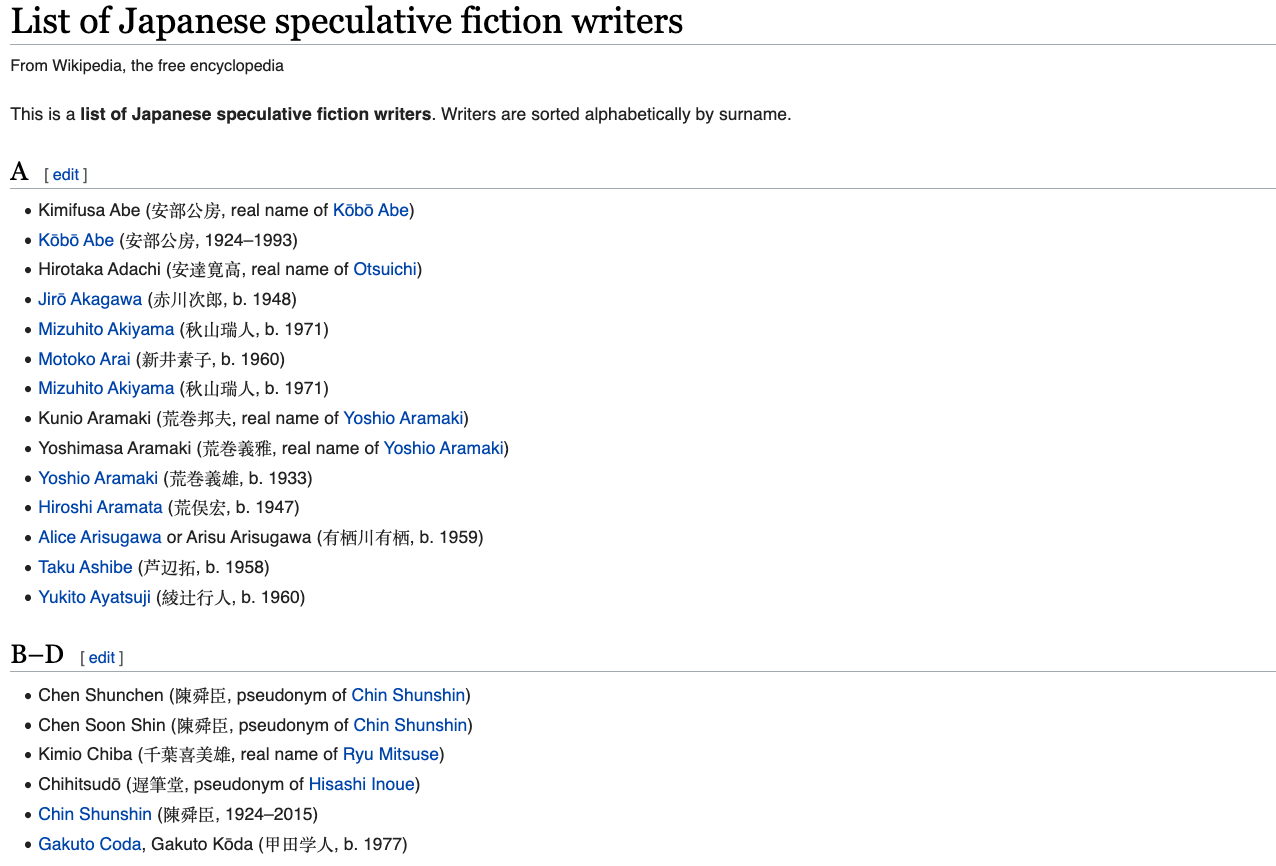}
    \caption{Excerpt of the Wikipedia page \texttt{List of Japanese speculative fiction writers} displaying the subjects in an \textbf{enumeration} layout.}
    \label{fig:list-page-enum}
\end{figure}

While Wikipedia's infoboxes and categories have been the subject of many information extraction efforts of knowledge graphs already, list pages have - despite their obvious wealth of information - received very little attention. For entities of the page \texttt{List of Japanese speculative fiction writers} (shown in Fig.~\ref{fig:list-page-enum}), we can derive several bits of information: \emph{(type, Writer)}, \emph{(nationality, Japan)}, and \emph{(genre, Speculative Fiction)}.

In contrast to finding entities of a category, finding such entities among all the entities mentioned in a list page is a non-trivial problem. We will refer to these entities, that are instances of the concept expressed by the list page, as its \emph{subject entities}. Unlike categories, list pages are an informal construct in Wikipedia. Hence, the identification of their subject entities brings up several challenges: While list pages are usually formatted as enumeration or table, they have no convention of how the information in them is structured. For example, subject entities can be listed somewhere in the middle of a table (instead of in the first column) and enumerations can have multiple levels. Furthermore, context information may not be available (it is difficult to find \emph{Japanese speculative fiction writers} in a list if one doesn't know to look for \emph{writers}).

In this paper, we introduce an approach for identifying subject entities in Wikipedia list pages and provide the following contributions in particular:
\begin{itemize}
    \item An approach for the construction of a combined taxonomy of Wikipedia categories, lists and DBpedia types.
    \item A distantly supervised machine learning approach for the extraction of subject entities from Wikipedia list pages.
    \item 700k new entities, 7.5M additional type statements, and 3.8M additional facts for DBpedia that are published as RDF triples and as a standalone knowledge graph called CaLiGraph\footnote{\url{http://caligraph.org}}.
\end{itemize}

The rest of this paper is structured as follows. Section~\ref{relatedwork} frames the approach
described in this paper in related works. Section~\ref{categories-and-lists-in-wikipedia} introduces the idea of entity extraction from list pages, followed by a description of our approach in Section~\ref{distantly-supervised-entity-extraction-from-list-pages}. In Section~\ref{results-and-discussion},
we discuss results and present an empirical evaluation of our approach. We close with a summary
and an outlook on future developments.

\section{Related Work} \label{relatedwork}
The extraction of knowledge from structured elements in Wikipedia is mostly focused on two fields: Firstly, the field of taxonomy induction, where most of the approaches use the category graph of Wikipedia to derive a taxonomy, and, secondly, the application of information extraction methods to derive facts from various (semi-)structured sources like infoboxes, tables, lists, or abstracts of Wikipedia pages.

The approach of Ponzetto and Navigli \cite{ponzetto2009large} was one of the first to derive a large taxonomy from Wikipedia categories by putting their focus on the lexical head of a category. They exploit the fact that almost exclusively categories with plural lexical heads are useful elements of a taxonomy. Hence, they are able to clean the category graph from non-taxonomic categories and relationships. Several other approaches create a combined taxonomy of the category graph and additional resources like WordNet (YAGO \cite{mahdisoltani2013yago3}) or Wikipedia pages (WiBi \cite{flati2014two}).

The distant supervision paradigm \cite{mintz2009distant} is used extensively for information extraction in Wikipedia as it provides an easy way to automatically gather large amounts of training data with a low error rate. Usually, some form of knowledge base is used as background knowledge to generate training data from a target corpus. In the original work, Mintz et al. use Freebase as background knowledge to extract information from Wikipedia. \cite{aprosio2013extending} extend this approach by using DBpedia as background knowledge.

Regarding list pages, Paulheim and Ponzetto \cite{paulheim2013extending} frame their general potential as a source of knowledge in Wikipedia. They propose to use a combination of statistical and NLP methods to extract knowledge and show that, by applying them to a single list page, they are able to extract a thousand new statements. \cite{kuhn2016type} infer types for entities on list pages and are thus most closely related to our approach. To identify subject entities of the list pages, they rely on information from DBpedia (e.g. how many relations exist between entities on the list page) and are consequently only able to infer new types for existing DBpedia entities. They use a score inspired by TF-IDF to find the type of a list page and are able to extract 303,934 types from 2,000 list pages with an estimated precision of 86.19\%.

Apart from list pages, entity and relation extraction in Wikipedia is applied to structured sources like infoboxes \cite{wu2007autonomously}, abstracts \cite{heist2017language,schrage2019extracting}, and tables \cite{bhagavatula2013methods,munoz2014using}.

With the exploitation of the structured nature of list pages to extract previously unknown entities as well as factual information about them, we see our approach as a useful addition to the existing literature where the focus is set primarily on enriching the ontology or adding information for existing entities.

\section{Categories and List Pages in Wikipedia} \label{categories-and-lists-in-wikipedia}
The Wikipedia Category Graph (WCG) has been used extensively for taxonomy induction (e.g. in \cite{flati2014two,mahdisoltani2013yago3}) and has proven to yield highly accurate results. The WCG has a subgraph consisting of list categories,\footnote{A list category is a Wikipedia category that starts with the prefix \emph{Lists of}.} which organizes many of the list page articles in Wikipedia. The list page \texttt{List of Japanese speculative fiction writers} (Fig.~\ref{fig:list-page-enum}), for example, is a member of the list category \texttt{Lists of Japanese writers}, which in turn has the parent list category \texttt{Lists of writers by nationality}, and so on.

As this subgraph is part of the WCG, we can use the list categories as a natural extension of a taxonomy induced by the WCG (e.g., by linking \texttt{Lists of Japanese writers} to the respective category \texttt{Japanese writers}). This comes with the benefit of including list pages into the taxonomy (i.e., we can infer that \texttt{List of Japanese speculative fiction writers} is a sub-concept of the category \texttt{Japanese writers}). Despite their obvious potential, neither list categories nor list pages have yet explicitly been used for taxonomy induction. 

In each list page, some of the links point to entities in the category the list page reflects, others do not. In the list page \texttt{List of Japanese speculative fiction writers}, for example, some links point to pages about such writers (i.e. to its subject entities), while others point to specific works by those writers. To distinguish those two cases, the unifying taxonomy is of immense value. Through the hierarchical relationships between categories and list pages, we can infer that if an entity is mentioned in both a list page \emph{and} a related category, it is very likely a subject entity of the list page. Consequently, if an entity is mentioned in the list page \texttt{List of Japanese speculative fiction writers} and is contained in the category \texttt{Japanese writers}, it is almost certainly a Japanese speculative fiction writer.

In the remainder of this section we provide necessary background information of the resources used in our approach.

\subsubsection{The Wikipedia Category Graph.}
In the version of October 2016\footnote{We use this version in order to be compatible with the (at the time of conducting the experiments) most recent release of DBpedia: \url{https://wiki.dbpedia.org/develop/datasets}.} the WCG consists of 1,475,015 categories that are arranged in a directed, but not acyclic graph. This graph does not only contain categories used for the categorisation of content pages, but also ones that are used for administrative purposes (e.g., the category \texttt{Wikipedia articles in need of updating}). Similar to \cite{heist2019uncovering}, we use only transitive subcategories of the category \texttt{Main topic classifications} while also getting rid of categories having one of the following keywords in their name: \emph{wikipedia, lists, template, stub}.

The resulting filtered set of categories $\mathcal{C}^F$ contains 1,091,405 categories that are connected by 2,635,718 subcategory edges. We denote the set of entities in a category \emph{c} with $\mathcal{E}_c$, the set of all types in DBpedia with $\mathcal{T}$ and the set of types of an entity \emph{e} with $\mathcal{T}_e$.

\subsubsection{The Wikipedia List Graph.}
The set of list categories $\mathcal{C}^L$ contains 7,297 list categories (e.g., \texttt{Lists of People}), connected by 10,245 subcategory edges (e.g., \texttt{Lists of Celebrities} being a subcategory of \texttt{Lists of People}). The set of list pages $\mathcal{L}$ contains 94,562 list pages. Out of those, 75,690 are contained in at least one category in $\mathcal{C}^F$ (e.g., \texttt{List of Internet Pioneers} is contained in the category \texttt{History of the Internet}), 70,099 are contained in at least one category in $\mathcal{C}^L$ (e.g., \texttt{List of Internet Pioneers} is contained in the category \texttt{Lists of Computer Scientists}), and 90,430 are contained in at least one of the two.\footnote{Note that $\mathcal{C}^F$ and $\mathcal{C}^L$ are disjoint as we exclude categories with the word \emph{lists} in $\mathcal{C}^F$.}

\subsubsection{The Anatomy of List Pages.}
List pages can be categorised into one of three possible layout types \cite{kuhn2016type}: 44,288 pages list entities in a bullet point-like \textbf{enumeration}. The list page \texttt{List of Japanese speculative fiction writers} in Fig.~\ref{fig:list-page-enum} lists the subject entities in an enumeration layout. In this case, the subject entities are most often mentioned at the beginning of an enumeration entry. As some exceptions on the page show, however, this is not always the case.

\begin{figure}[t]
    \centering
    \includegraphics[width=\textwidth]{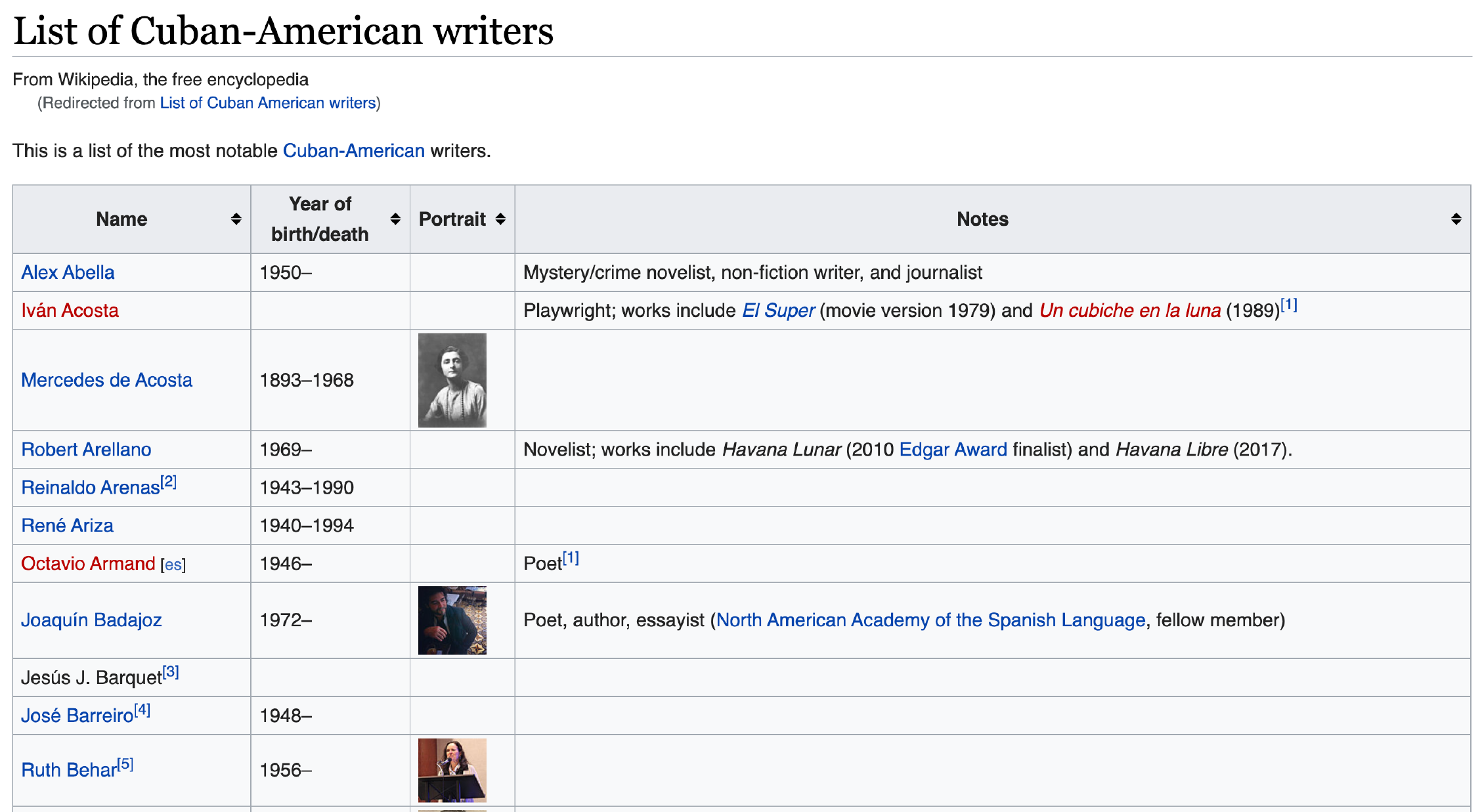}
    \caption{Excerpt of the Wikipedia page \texttt{List of Cuban-American writers} displaying the subjects in a \textbf{table} layout.}
    \label{fig:list-page-table}
\end{figure}

46,160 pages list entities in a \textbf{table} layout. An example of this layout is given in Fig.~\ref{fig:list-page-table}, where an excerpt of the page \texttt{List of Cuban-American writers} is shown. The respective subjects of the rows are listed in the first column, but this can also vary between list pages.

The remaining 4,114 pages do not have a consistent layout and are thus categorised as \textbf{undefined}\footnote{We heuristically label a list page as having one of the three layout types by looking for the most frequent elements: enumeration entries, table rows, or none of them.}. As our approach significantly relies on the structured nature of a list page, we exclude list pages with an undefined layout from our extraction pipeline.

For a list page \emph{l}, we define the task of identifying its subject entities $\mathcal{E}_l$ among all the mentioned entities $\mathcal{\widehat{E}}_l$ in \emph{l} as a binary classification problem. A mentioned entity is either classified as being a subject entity of \emph{l} or not. If not, it is usually mentioned in the context of an entity in $\mathcal{E}_l$ or for organisational purposes (e.g. in a \emph{See also} section). Looking at Figures~\ref{fig:list-page-enum}~and~\ref{fig:list-page-table}, mentioned entities are marked in blue (indicating that they have an own Wikipedia page and are thus contained in DBpedia) and in red (indicating that they do not have a Wikipedia page and are consequently no entities in DBpedia). Additionally, we could include possible entities that are not tagged as such in Wikipedia (e.g. \emph{Jesús J. Barquet} in the first column of Fig.~\ref{fig:list-page-table}) but we leave this for future work as it introduces additional complexity to the task. Of the three types of possible entities, the latter two are the most interesting as they would add the most amount of information to DBpedia. But it is also beneficial to identify entities that are already contained in DBpedia because we might be able to derive additional information about them through the list page they are mentioned in.

Note that for both layout types, \textbf{enumeration} and \textbf{table}, we find at most one subject entity per enumeration entry or table row. We inspected a subset of $\mathcal{L}$ and found this pattern to occur in every one of them.

\subsubsection{Learning Category Axioms with Cat2Ax}
The approach presented in this paper uses axioms over categories to derive a taxonomy from the category graph. Cat2Ax \cite{heist2019uncovering} is an approach that derives two kinds of axioms from Wikipedia categories: type axioms (e.g., for the category \texttt{Japanese writers} it learns that all entities in this category are of the type \emph{Writer}), and relation axioms (e.g., for the same category it learns that all entities have the relation \emph{(nationality, Japan)}). The authors use statistical and linguistic signals to derive the axioms and report a correctness of 96\% for the derived axioms.

\section{Distantly Supervised Entity Extraction from List Pages} \label{distantly-supervised-entity-extraction-from-list-pages}
\begin{figure}[t]
    \hspace*{-6mm}
    \centering
    \includegraphics[width=.95\textwidth]{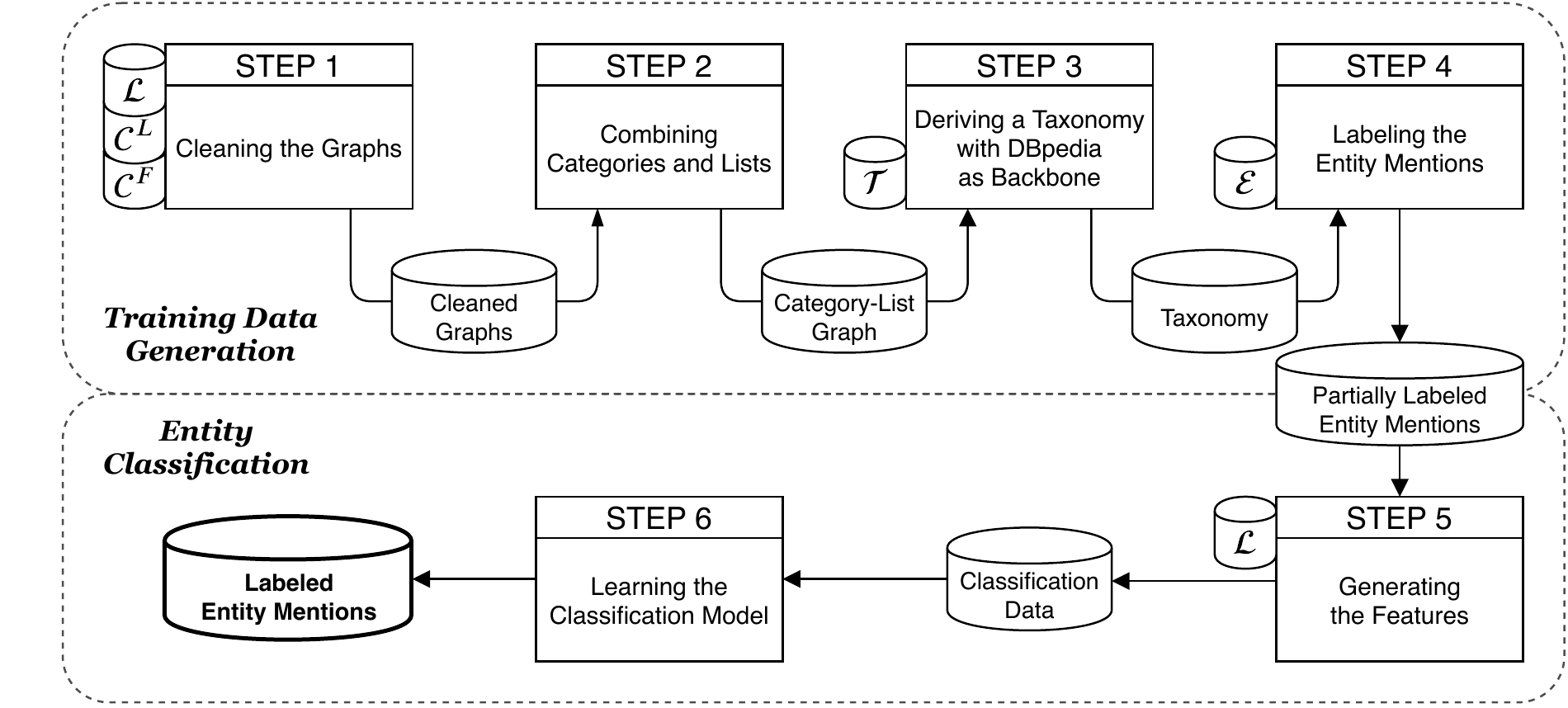}
    \caption{Overview of the pipeline for the retrieval of subject entities from list pages. Small cylindrical shapes next to a step indicate the use of external data, and large cylindrical shapes contain data that is passed between pipeline steps.}
    \label{fig:extraction-overview}
\end{figure}

The processing pipeline for the retrieval of subject entities from list pages in $\mathcal{L}$ is summarized in Fig.~\ref{fig:extraction-overview}. The pipeline consists of two main components: In the \textbf{Training Data Generation} we create a unified taxonomy of categories, lists, and DBpedia types. With distant supervision we induce positive and negative labels from the taxonomy for a part of the mentioned entities of list pages.

The resulting training data is passed to the \textbf{Entity Classification} component. There we enrich it with features extracted from the list pages and learn classification models to finally identify the subject entities.

\subsection{Training Data Generation} \label{training-data-generation}

\subsubsection{Step 1: Cleaning the Graphs\newline}
\begin{wrapfigure}{r}{0.5\textwidth}\centering
    \includegraphics[scale=0.37]{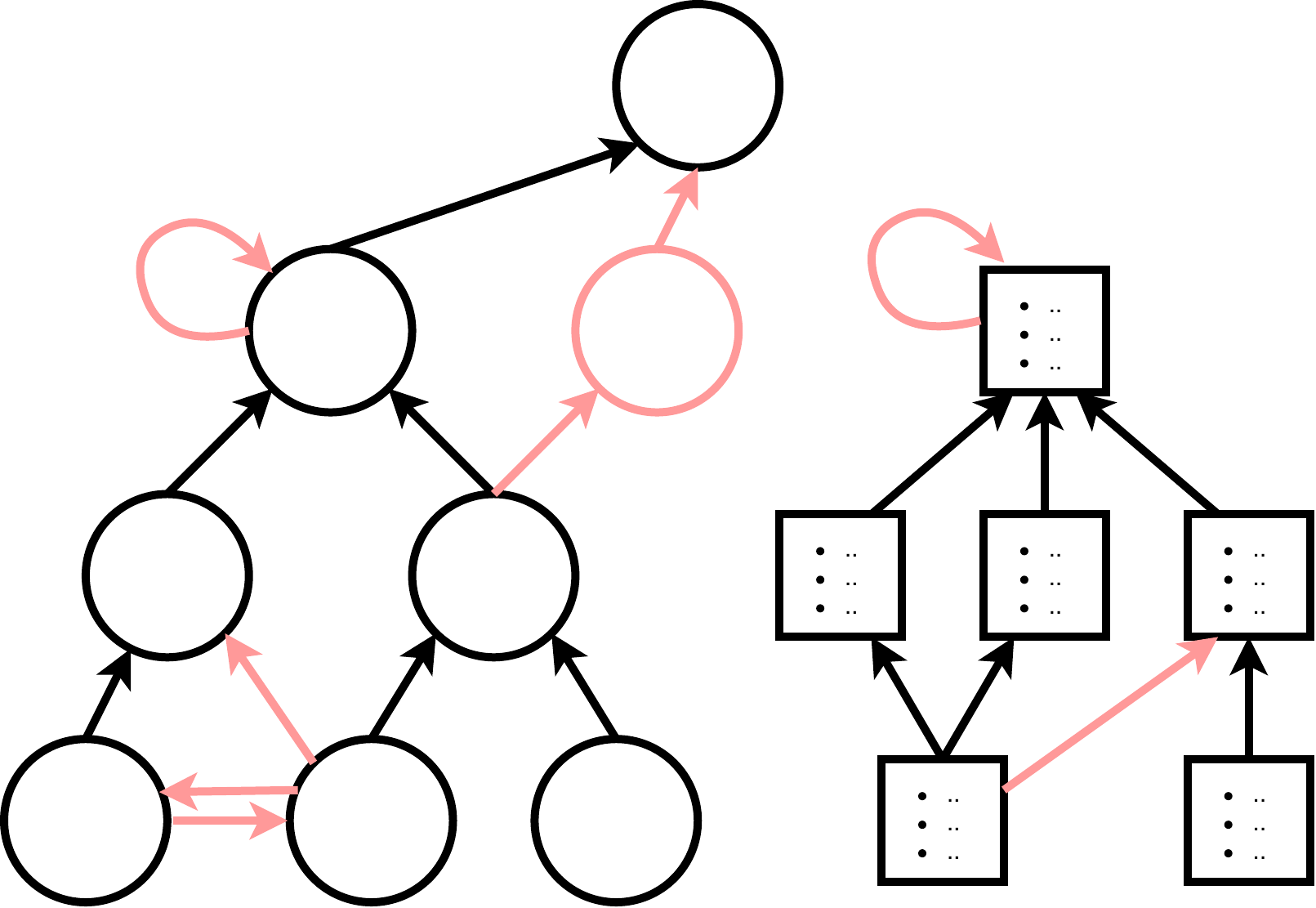}
    \caption{Possible invalid nodes and edges (marked in red) in the category graph (circles) and list graph (rectangles).}
    \label{fig:cleaning-the-graphs}
\end{wrapfigure}

\begin{figure}[t]
    \centering
    \includegraphics[width=.95\textwidth]{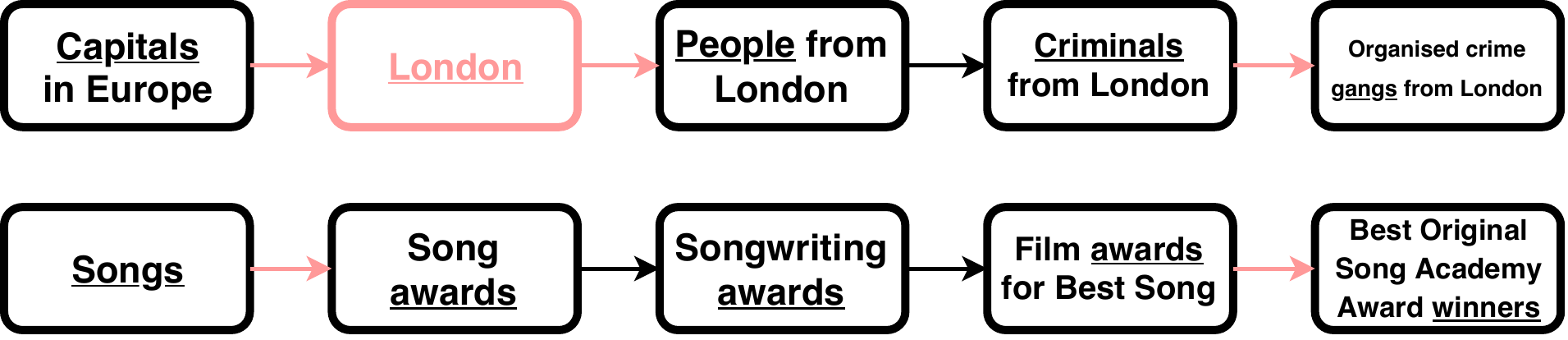}
    \caption{Examples of non-taxonomic nodes and edges (marked in red) that must be removed from the respective category graph or list graph.}
    \label{fig:cleaning-the-graphs-example}
\end{figure}

The initial category graph ($\mathcal{C}^F$ as nodes, subcategory relations as edges) and the initial list graph ($\mathcal{C}^L$ and $\mathcal{L}$ as nodes, subcategory relations and category membership as edges) both contain nodes and edges that have to be removed in order to convert them into valid taxonomies. Potential problems are shown in an abstract form in Fig.~\ref{fig:cleaning-the-graphs} and on an example in Fig.~\ref{fig:cleaning-the-graphs-example}. In particular, we have to remove nodes that do not represent proper taxonomic types (e.g. \texttt{London} in Fig.~\ref{fig:cleaning-the-graphs-example}). Additionally, we have to remove edges that either do not express a valid subtype relation (e.g. the edge from \texttt{Songs} to \texttt{Song awards} in Fig.~\ref{fig:cleaning-the-graphs-example}), or create cycles (e.g. the self-references in Fig.~\ref{fig:cleaning-the-graphs}).

For the removal of non-taxonomic nodes we rely on the observation made by \cite{ponzetto2009large}, that a Wikipedia category is a valid type in a taxonomy if its head noun is in plural. Consequently, we identify the head nouns of the nodes in the graph and remove all nodes with singular head nouns.\footnote{We use spaCy (\url{http://spacy.io}) for head noun tagging.}

For the removal of invalid edges we first apply a domain-specific heuristic to get rid of non-taxonomic edges and subsequently apply a graph-based heuristic that removes cycles in the graphs. The domain-specific heuristic is based on \cite{ponzetto2009large}: An edge is removed if the head noun of the parent is not a synonym or a hypernym of the child's head noun. In Fig.~\ref{fig:cleaning-the-graphs-example} the head nouns of nodes are underlined; we remove, for example, the edge from \texttt{Songs} to \texttt{Song awards} as the word \emph{songs} is neither a synonym nor a hypernym of \emph{awards}.

We base our decision of synonym and hypernym relationships on a majority vote from three sources: (1) We parse the corpus of Wikipedia for Hearst patterns \cite{hearst1992automatic}.\footnote{Patterns that indicate a taxonomic relationship between two words like "X is a Y".} (2) We extract them from WebIsALOD \cite{hertling2017webisalod}, a large database of hypernyms crawled from the Web. (3) We extract them directly from categories in Wikipedia. To that end, we apply the Cat2Ax approach \cite{heist2019uncovering} which computes robust type and relation axioms for Wikipedia categories from linguistic and statistical signals. For every edge in the category graph, we extract a hypernym relationship between the head noun of the parent and the head noun of the child if we found matching axioms for both parent and child. E.g., if we find the axiom that every entity in the category \texttt{People from London} has the DBpedia type \emph{Person} and we find the same axiom for \texttt{Criminals from London}, then we extract a hypernym relation between \emph{People} and \emph{Criminals}.

As a graph-based heuristic to resolve cycles, we detect edges that are part of a cycle and remove the ones that are pointing from a deeper node to a higher node in the graph.\footnote{We define the depth of a node in the graph as the length of its shortest path to the root node \texttt{Main topic classifications}.} If cycles can not be resolved because edges point between nodes on the same depth level, those are removed as well.

Through the cleaning procedure we reduce the size of the category graph from 1,091,405 nodes and 2,635,718 edges to 738,011 nodes and 1,324,894 edges, and we reduce the size of the list graph from 77,396 nodes and 105,761 edges to 77,396 nodes and 95,985 edges.

\subsubsection{Step 2: Combining Categories and Lists\newline}

\begin{wrapfigure}{r}{0.5\textwidth}\centering
    \includegraphics[scale=0.37]{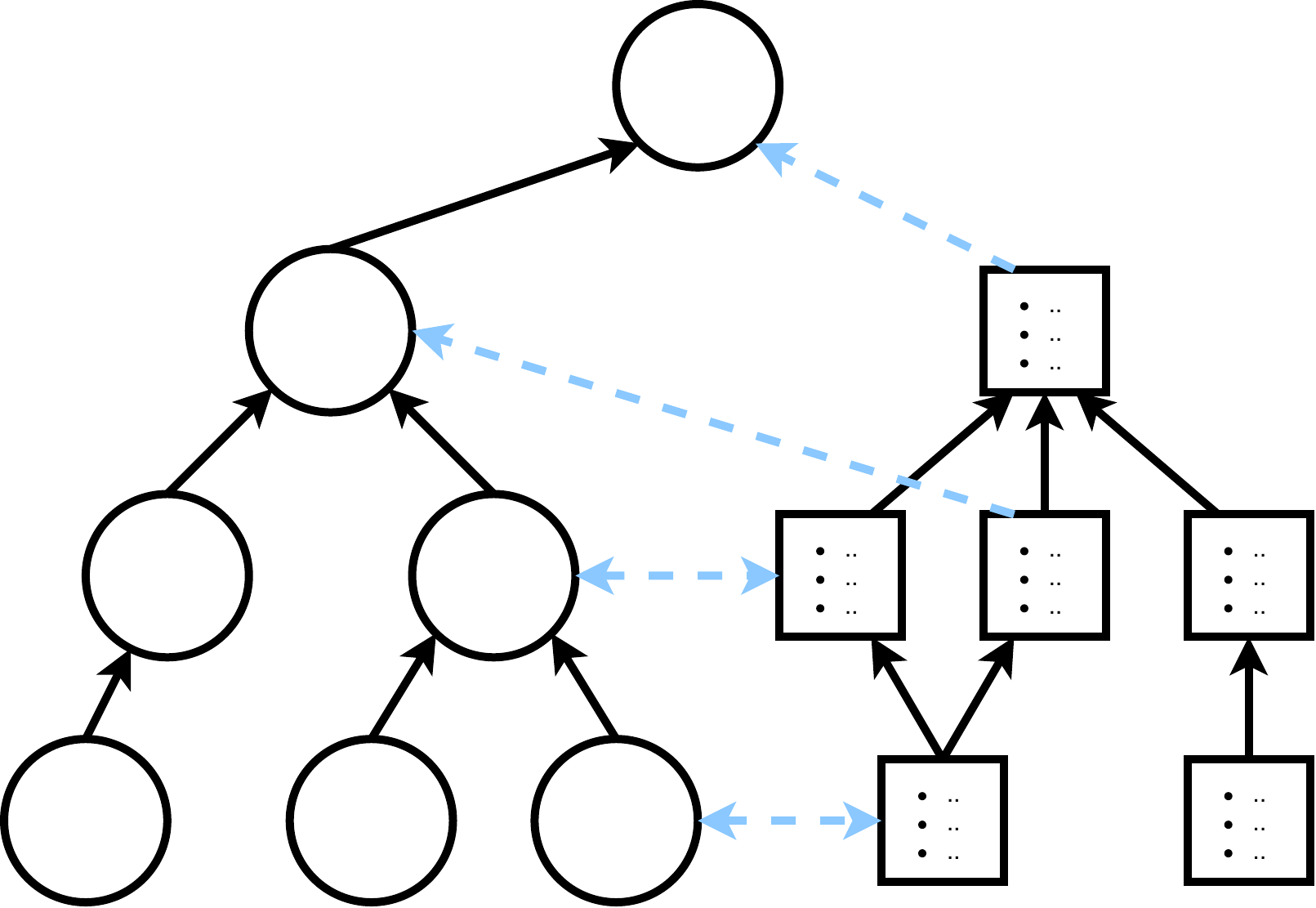}
    \caption{Possible connections between the category graph and the list graph.}
    \label{fig:combining-categories-and-lists}
\end{wrapfigure}

For a combined taxonomy of categories and lists, we find links between lists and categories based on linguistic similarity and existing connections in Wikipedia. As Fig.~\ref{fig:combining-categories-and-lists} shows, we find two types of links: equivalence links and hypernym links. We identify the former by looking for category-list pairs that are either named similar (e.g. \texttt{Japanese writers} and \texttt{Lists of Japanese writers}) or are synonyms (e.g. \texttt{Media in Kuwait} and \texttt{Lists of Kuwaiti media}). With this method we find 24,383 links.

We extract a hypernym link (similar to the method that we applied in Step 1) if the head noun of a category is a synonym or hypernym of a list's head noun. However, we limit the candidate links to existing edges in Wikipedia (i.e. the subcategory relation between a list category and a category, or the membership relation between a list page and a category) in order to avoid false positives. With this method we find 19,015 hypernym links. By integrating the extracted links into the two graphs, we create a category-list graph with 815,543 nodes (738,011 categories, 7,416 list categories, 70,116 list pages) and 1,463,423 edges.

\subsubsection{Step 3: Deriving a Taxonomy with DBpedia as Backbone\newline}
\begin{wrapfigure}{r}{0.5\textwidth}\centering
    \vspace*{-8.6mm}
    \includegraphics[scale=0.37]{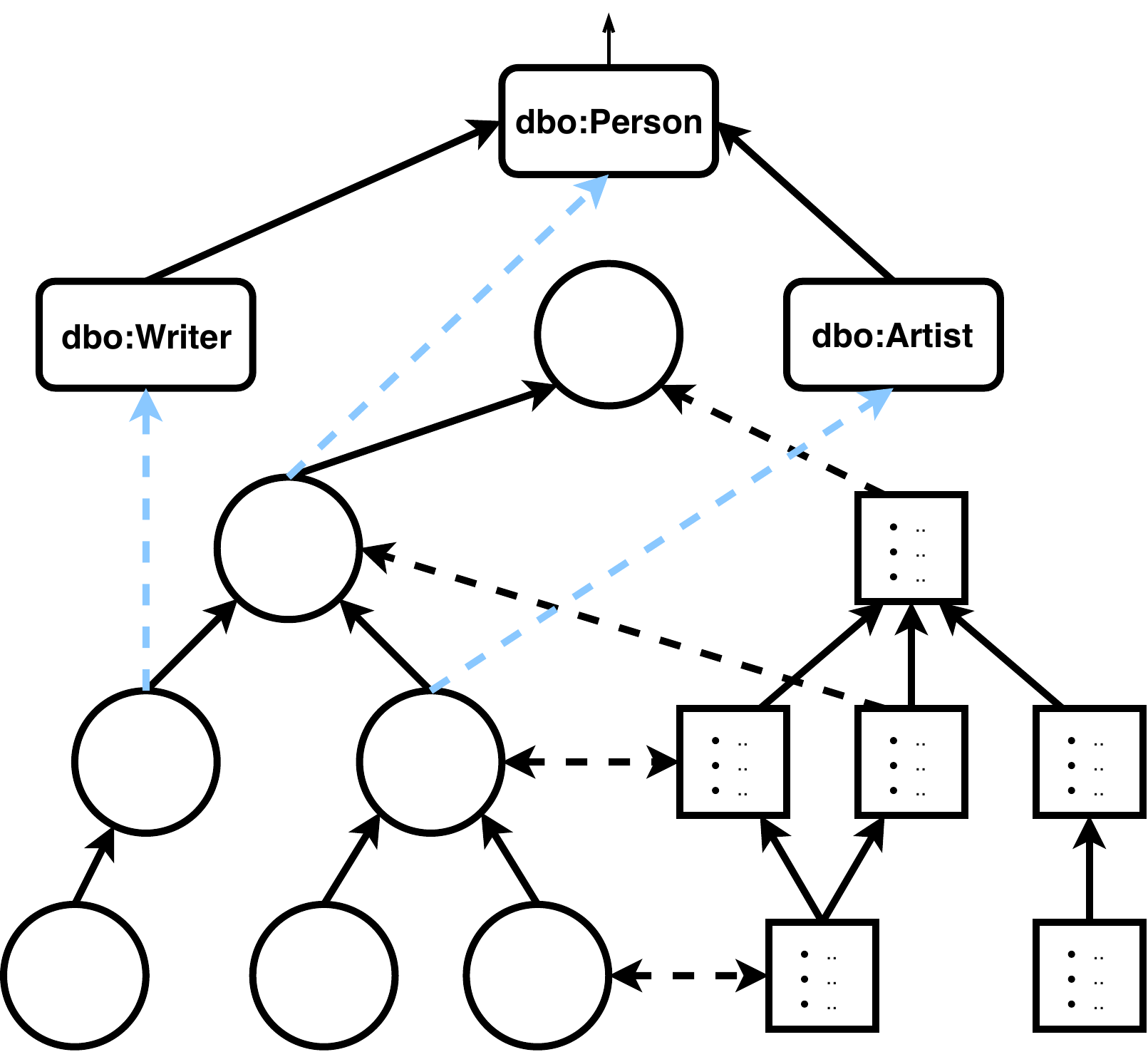}
    \caption{Extension of the category-list taxonomy with DBpedia as a backbone.}
    \label{fig:deriving-a-taxonomy}
\end{wrapfigure}

As a final step, we connect the category-list graph with the DBpedia taxonomy (as depicted in Fig.~\ref{fig:deriving-a-taxonomy}). To achieve that, we again apply the Cat2Ax approach to our current graph to produce type axioms for the graph nodes. E.g., we discover the axiom that every entity in the category \texttt{Japanese writers} has the DBpedia type \emph{Writer}, thus we use the type as a parent of \texttt{Japanese writers}. Taking the transitivity of the taxonomy into account, we find a DBpedia supertype for 88\% of the graph's nodes.

\subsubsection{\mbox{Step 4: Labeling the Entity Mentions\newline}}
\makebox[\textwidth][r]{We parse all entity mentions in list pages directly from the Wikitext using the} dumps provided by DBpedia and using WikiTextParser\footnote{https://github.com/5j9/wikitextparser} as a markup parser.

We compute the training data for mentioned entities $\mathcal{\widehat{E}}_l$ of a list page \emph{l} directly from the taxonomy. To that end, we define two mapping functions:
\begin{eqnarray}
related: \mathcal{L} \rightarrow P(\mathcal{C}^F) \\
types: \mathcal{L} \rightarrow P(\mathcal{T})
\label{eq:example_axioms}
\end{eqnarray}

The function \emph{related(l)} from Definition~1 returns the subset of $\mathcal{C}^F$ that contains the taxonomically equivalent or most closely related categories for \emph{l}. For example, \emph{related(}\texttt{List of Japanese speculative fiction writers}\emph{)} returns the category \texttt{Japanese writers} and all its transitive subcategories (e.g. \texttt{Japanese women writers}). To find \emph{related(l)} of a list page \emph{l}, we traverse the taxonomy upwards starting from \emph{l} until we find a category \emph{c} that is contained in $\mathcal{C}^F$. Then we return \emph{c} and all of its children.

With this mapping, we assign positive labels to entity mentions in \emph{l}, if they are contained in a category in \emph{related(l)}:

\begin{equation}\label{eq:positive-labels}
    \mathcal{\widehat{E}}^+_l = \left\{e | e \in \mathcal{\widehat{E}}_l \land \exists c \in related(l) : e \in \mathcal{E}_c  \right\}
\end{equation}

In the case of \texttt{List of Japanese speculative fiction writers}, $\mathcal{\widehat{E}}^+_l$ contains all entities that are mentioned on the list page \emph{and} are members of the category \texttt{Japanese writers} or one of its subcategories.

The function \emph{types(l)} from Definition~2 returns the subset of the DBpedia types $\mathcal{T}$ that best describes entities in \emph{l}. For example, \emph{types(}\texttt{List of Japanese speculative fiction writers}\emph{)} returns the DBpedia types \emph{Agent}, \emph{Person}, and \emph{Writer}. To find \emph{types(l)}, we retrieve all ancestors of \emph{l} in the taxonomy and return those contained in $\mathcal{T}$.

With this mapping, we assign a negative label to an entity \emph{e} mentioned in \emph{l}, if there are types in $\mathcal{T}_e$ that are disjoint with types in \emph{types(l)}:

\begin{equation}\label{eq:negative-labels}
    \mathcal{\widehat{E}}^-_l = \left\{e | e \in \mathcal{\widehat{E}}_l \land \exists t_e \in \mathcal{T}_e, \exists t_l \in types(l) : disjoint(t_e,t_l) \right\}
\end{equation}

To identify disjointnesses in Eq.~\ref{eq:negative-labels}, we use the disjointness axioms provided by DBpedia as well as additional ones that are computed by the methods described in \cite{topper2012dbpedia}. DBpedia declares, for example, the types \emph{Person} and \emph{Building} as disjoint, and the type \emph{Person} is contained in \emph{types(}\texttt{List of Japanese speculative fiction writers}\emph{)}. Consequently, we label any mentions of buildings in the list page as negative examples.

In addition to the negative entity mentions that we retrieve via Eq.~\ref{eq:negative-labels}, we label entities as negative using the observation we have made in Section~\ref{categories-and-lists-in-wikipedia}: As soon as we find a positive entity mention in an enumeration entry or table row, we label all the remaining entity mentions in that entry or row as negative.

For enumeration list pages, we find a total of 9.6M entity mentions. Of those we label 1.4M as positive and 1.4M as negative. For table list pages, we find a total of 11M entity mentions. Of those we label 850k as positive and 3M as negative.

\subsection{Entity Classification} \label{entity-extraction}

\subsubsection{Step 5: Generating the Features\newline}
{\renewcommand{\arraystretch}{1.25}%
	\begin{table}[t]
		\centering
		\begin{tabular}{| C{0.5cm} | C{2.5cm} | L{8.8cm} |}
			\hline
			& \textbf{Feature Type} & \multicolumn{1}{c|}{\textbf{Features}} \\
			\hline \hline
			 \parbox[t]{2mm}{\multirow{3}{*}{\rotatebox[origin=c]{90}{\textbf{Shared}}}} & Page & \# sections \\
			 & Positional & Position of section in LP \\
			 & Linguistic & Section title, POS/NE tag of entity and its direct context \\
			\hline \hline
			 \parbox[t]{2mm}{\multirow{5}{*}{\rotatebox[origin=c]{90}{\textbf{Enum}}}} & Page & \# entries, Avg. entry indentation level, Avg. entities/ words/characters per entry, Avg. position of first entity \\
			 & Positional & Position of entry in enumeration, Indentation level of entry, \# of sub-entries of entry, Position of entity in entry \\
			 & Custom & \# entities in current entry, \# mentions of entity in same/other enumeration of LP \\
			\hline \hline
			 \parbox[t]{2mm}{\multirow{6}{*}{\rotatebox[origin=c]{90}{\textbf{Table}}}} & Page & \# tables, \# rows, \# columns, Avg. rows/columns per table, Avg. entities/words/characters per row/column, Avg. first column with entity \\
			 & Positional & Position of table in LP, Position of row/column in table, Position of entity in row  \\
			 & Linguistic & Column header is synonym/hyponym of word in LP title \\
			 & Custom & \# entities in current row, \# mentions of current entity in same/other table of LP \\
			\hline
		\end{tabular}
		\caption{Features of the machine learning model grouped by list page type and by feature type. Page features are computed for the complete list page (LP) and do not vary between entity mentions. For page features, we include standard deviations as features in addition to averages.}
		\label{tab:features}
	\end{table}
}

For a single data point (i.e. the mention of an entity in a specific list page), we generate a set of features that is shown in Table~\ref{tab:features}. Shared features are created for entity mentions of both enumeration and table list pages.

Features of the type \emph{Page} encode characteristics of the list page and are hence similar for all entity mentions of the particular page. Features of the type \emph{Positional, Linguistic, Custom} describe the characteristics of a single entity mention and its immediate context.

\subsubsection{Step 6: Learning the Classification Model\newline}
{\renewcommand{\arraystretch}{1.25}%
	\begin{table}[t]
		\centering
		\begin{tabular}{| C{5.3cm} | C{1cm} | C{1cm} | C{1cm} | C{1cm} | C{1cm} | C{1cm} |}
			\hline
			 & \multicolumn{3}{c|}{\textbf{Enum}} & \multicolumn{3}{c|}{\textbf{Table}} \\
			\hline
			\textbf{Algorithm} & P & R & F1 & P & R & F1 \\
			\hline \hline
			Baseline (pick first entity) & 74 & 96 & 84 & 64 & 53 & 58 \\
			\hline \hline
			Naive Bayes & 80 & 90 & 84 &  34 & \textbf{91} & 50 \\
			\hline
			Decision Tree & 82 & 78 & 80 &  67 & 66 & 67 \\
			\hline
			Random Forest & 85 & \textbf{90} & \textbf{87} &  85 & 71 & \textbf{77} \\
			\hline
			XG-Boost & \textbf{90} & 83 & 86 &  \textbf{90} & 53 & 67 \\
			\hline
			Neural Network (MLP) & 86 & 84 & 85 &  78 & 72 & 75 \\
			\hline
			SVM & 86 & 60 & 71 &  73 & 33 & 45 \\
			\hline
		\end{tabular}
		\caption{Precision (P), Recall (R), and F-measure (F1) in percent for the positive class (i.e. true subject entities in a list page) of various classification models.}
		\label{tab:classification-results}
	\end{table}
}

To find a suitable classification model, we conduct an initial experiment on six classifiers (shown in Table~\ref{tab:classification-results}) and compare them with the obvious baseline of always picking the first entity mention in an enumeration entry or table row. We compute the performance using 10-fold cross validation while taking care that all entity mentions of a list page are in the same fold. In each fold, we use 80\% of the data for training and 20\% for validation. For all the mentioned classifiers, we report their performances after tuning their most important parameters with a coarse-grained grid search.

Table~\ref{tab:classification-results} shows that all applied approaches outperform the baseline in terms of precision. The XG-Boost algorithm scores highest in terms of precision while maintaining rather high levels of recall. Since we want to identify entities in list pages with highest possible precision, we use the XG-Boost model. After a fine-grained parameter tuning, we train models with a precision of 91\% and 90\%, and a recall of 82\% and 55\% for enumeration and table list pages, respectively.\footnote{The models are trained using the scikit-learn library: \url{https://scikit-learn.org/}.} Here, we split the dataset into 60\% training, 20\% validation, and 20\% test data.

\section{Results and Discussion} \label{results-and-discussion}
\subsubsection{Entities.}
\begin{figure}
\centering
\begin{minipage}{.48\textwidth}
    \centering
    \includegraphics[width=\textwidth]{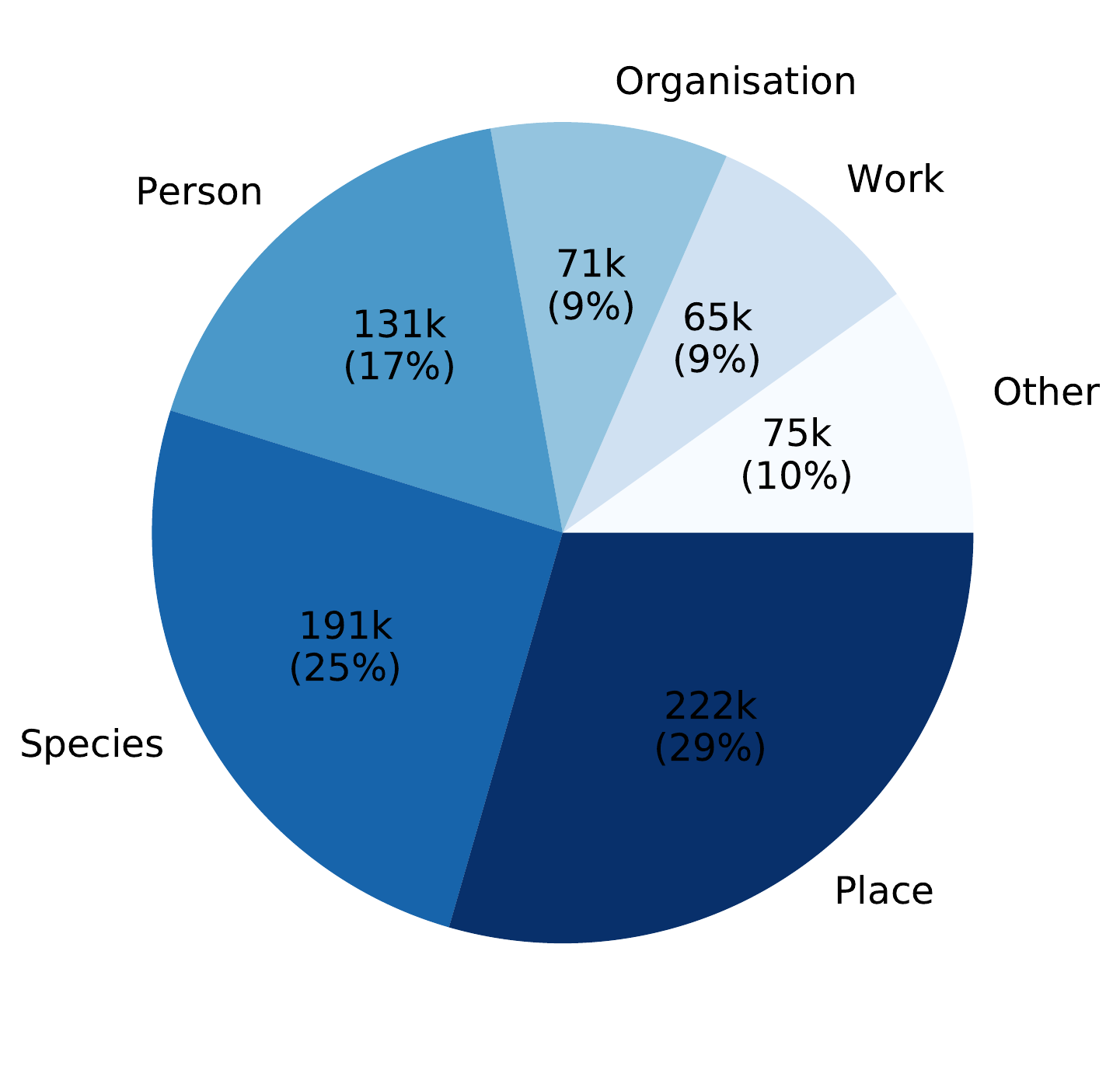}
    \caption{Distribution of entities that are added to DBpedia based on high-level types.}
    \label{fig:new-instance-types}
\end{minipage}
\hspace{2mm}
\begin{minipage}{.48\textwidth}
    \centering
    \vspace{9mm}
    \includegraphics[width=\textwidth]{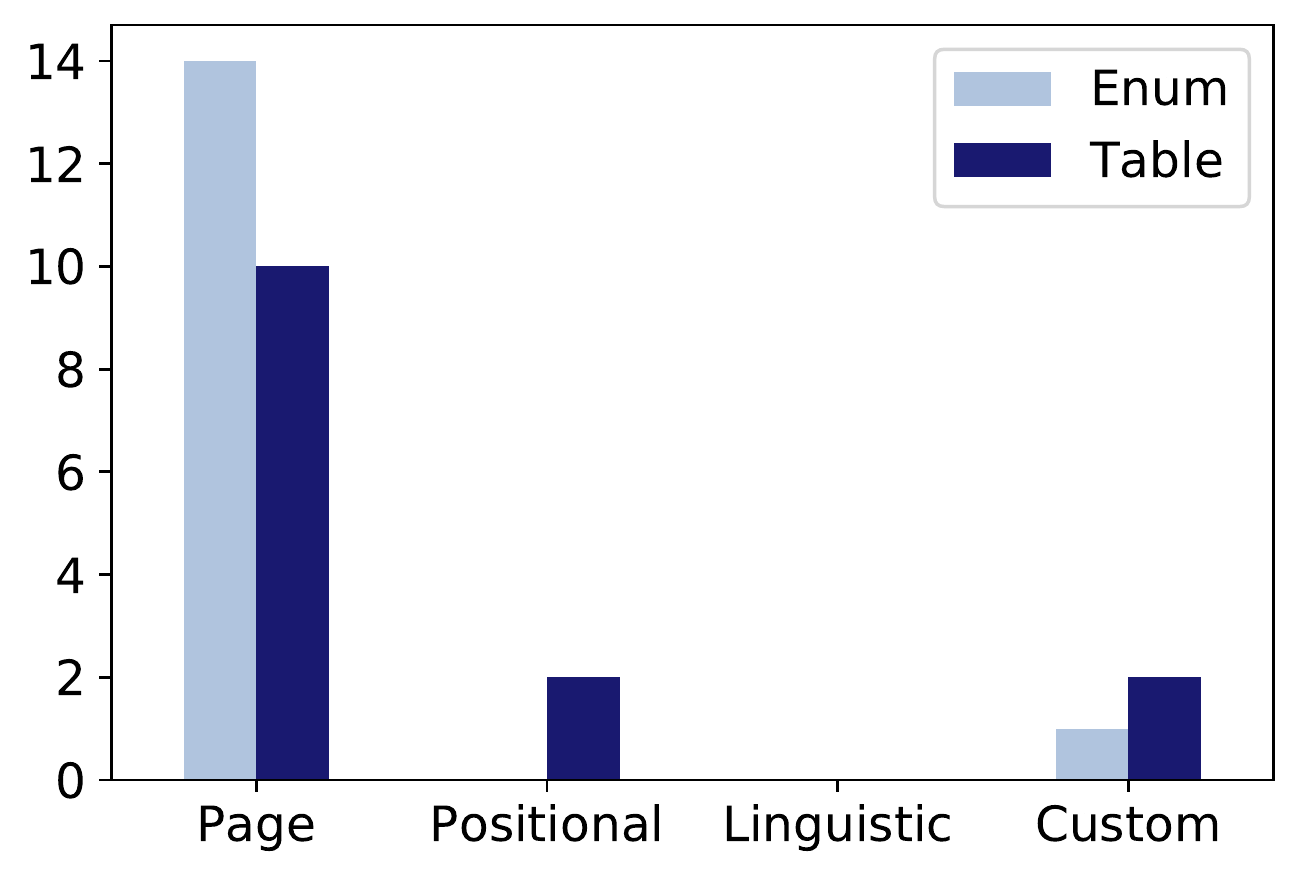}
    \vspace{3mm}
    \caption{The 15 most important features used by XG-Boost grouped by respective feature type.}
    \label{fig:important-features}
\end{minipage}
\end{figure}

In total, we extract 1,549,893 subject entities that exist in DBpedia already. On average, an entity is extracted from 1.86 different list pages. Furthermore, we extract 754,115 subject entities that are new to DBpedia (from 1.07 list pages on average). Based on the list pages they have been extracted from, we assign them DBpedia types (i.e., the supertypes of the list page in the derived taxonomy). Fig.~\ref{fig:new-instance-types} shows the distribution of new entities over various high-level types.

\subsubsection{Entity Types.}
\begin{figure}[t]
    \centering
    \includegraphics[width=.9\textwidth]{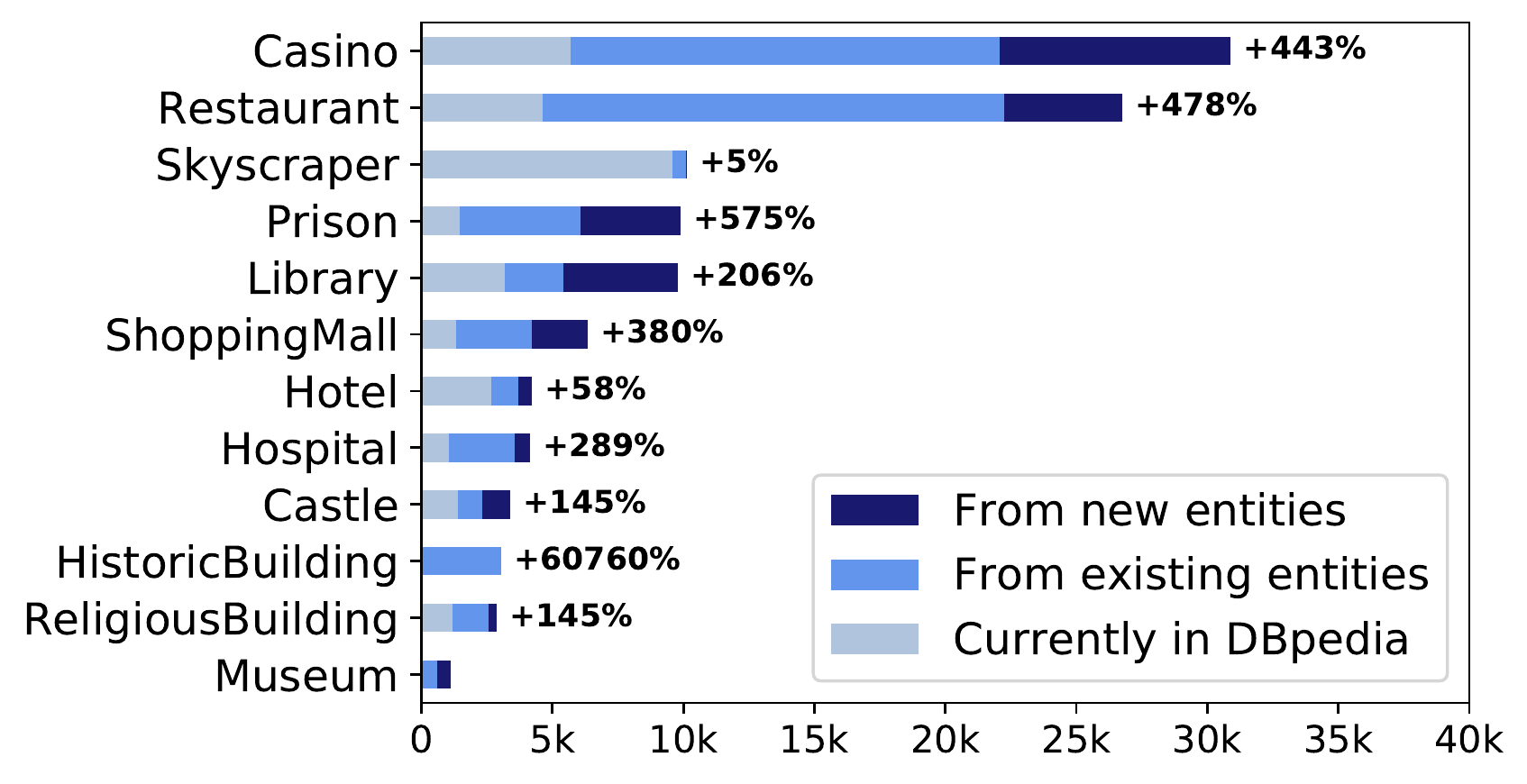}
    \caption{Comparison of the number of type statements that are currently in DBpedia with additional statements found by our approach for all subtypes of the DBpedia type \emph{Building}.}
    \label{fig:type-increase}
\end{figure}

Overall, we generate 7.5M new type statements for DBpedia: 4.9M for entities in DBpedia (we assign a type to 2M previously untyped entities), and 2.6M for new entities (we find an average of 3.5 types per new entity). This is an increase of 51.15\% in DBpedia's total type statements. We especially generate statements for types that are rather specific, i.e., deep in the ontology.\footnote{We define the depth of a type in DBpedia as the length of its shortest path to the root type \emph{owl:Thing}.} Adding all the generated type statements to DBpedia, the average type depth increases from 2.9 to 2.93. For new entities, we have an average type depth of 3.06. Fig.~\ref{fig:type-increase} shows the increase of type statements for the subtypes of the DBpedia type \emph{Building}. For almost all of them, we increase the amount of type statements by several orders of magnitude.

\subsubsection{Entity Facts.}
Besides type statements, we also infer relational statements using the relation axioms that we generated via Cat2Ax. In total, we generate 3.8M relational statements: 3.3M for existing entities in DBpedia and 0.5M for new entities. For some previously unknown entities we discover quite large amounts of facts. For the moth species \emph{Rioja}\footnote{\url{http://caligraph.org/resource/Rioja_(moth)}}, for example, we discover the type \emph{Insect} and information about its \emph{class, family, order,} and \emph{phylum}. For \emph{Dan Stulbach}\footnote{\url{http://caligraph.org/resource/Dan_Stulbach}} we discover the type \emph{Person} and information about his \emph{birth place, occupation}, and \emph{alma mater}.

\subsubsection{Evaluation.}
We evaluate the correctness of both the constructed taxonomy and the inferred statements.

To validate the taxonomy we conducted an evaluation on the crowd-sourcing platform Amazon MTurk.\footnote{\url{https://mturk.com}} We randomly sampled 2,000 edges of the taxonomy graph and asked three annotators each whether the edge is taxonomically correct. The edges have been evaluated as correct in 96.25\% ($\pm$0.86\%) of the cases using majority vote (with an inter-annotator agreement of 0.66 according to Fleiss' kappa \cite{fleiss_kappa}).

The correctness of the inferred type and relation statements are strongly dependent on the Cat2Ax approach as we use its axioms to generate the statements. The authors of Cat2Ax report a correctness of 96.8\% for type axioms and 95.6\% for relation axioms. For the resulting type and relation statements (after applying the axioms to the entities of the categories) they report a correctness of 90.8\% and 87.2\%, respectively. However, the original Cat2Ax approach does not rely on a complete taxonomy of categories but computes axioms for individual categories without considering hierarchical relationships between them. In contrast, we include information about the subcategories of a given category while generating the axioms. An inspection of 1,000 statements\footnote{We inspect 250 type and relation statements for both existing and new entities.} by the authors yields a correctness of 99\% ($\pm$1.2\%) for existing and 98\% ($\pm$1.7\%) for new type statements, and 95\% ($\pm$2.7\%) for existing and 97\% ($\pm$2.1\%) for new relation statements.

\subsubsection{Classification Models.}
With values of 91\% and 90\% the precision of the classification models is significantly lower than the correctness of the extracted type and relation statements. At first glance this is a contradiction because, although the models extract entities and not statements, a statement is obviously incorrect if it has been created for the wrong entity. But we have to consider that the training data, which was used to train and evaluate the models, has been created using distant supervision. Hence, it is likely to contain errors (e.g. due to wrong inheritance relations in the taxonomy). The fact that the final output of the processing pipeline has a higher correctness than the evaluation results of the models imply, indicates that the models are in fact able to learn meaningful patterns from the training data.

Fig.~\ref{fig:important-features} shows the feature types of the 15 features that have the highest influence on the decision of the final XG-Boost models. Almost all of them are features of the type \emph{Page}, i.e. features that describe the general shape of the list page the entities are extracted from. Features of the other types, that describe the immediate context of an entity, are used only very sparsely. This might be an indicator that, to bridge the gap in recall between the classification models and the baseline, we have to develop models that can make better use of the structure of a list page. Accordingly, we see the biggest potential in an adapted machine learning approach that, instead of classifying every entity mention in isolation, uses a holistic perspective and identifies the set of mentions that fit the list page best, given its structure.

\section{Conclusion} \label{conclusion}
In this paper we have presented an approach for the extraction of entities from Wikipedia list pages in order to enrich DBpedia with additional entities, type statements, and facts. We have shown that by creating a combined taxonomy from the WCG, its subgraph formed of lists, and DBpedia, we are able to train highly precise entity extraction models using distant supervision.

To extend our approach, we investigate in two directions. Firstly, we want to further improve the entity extraction by considering entities that are not explicitly tagged as such in list pages. In alignment with that we are developing a method to extract entities of a list page based on a joint likelihood instead of evaluating each entity mention in isolation. To that end we are experimenting with additional features that take the visual layout of a list page and alignment of entities into account.

As soon as we include untagged entities in the extraction, we will have to develop an entity disambiguation mechanism in order to create separate entities for homonyms. For this, we expect the distance between entities in the taxonomy to be a helpful indicator. 

Secondly, we investigate an application of our extraction approach to any kind of structured markup in Wikipedia (e.g. enumerations and tables that occur anywhere in Wikipedia), and, ultimately, to markup of arbitrary pages on the web. To achieve that, we want to combine the information about entity alignment on the web page with the available semantic information as outlined in \cite{heist2018towards}.
\\ \\
Code and results of this paper are published on \url{http://caligraph.org}.

\bibliographystyle{splncs04}
\bibliography{references}

\begin{thebibliography}{10}
\providecommand{\url}[1]{\texttt{#1}}
\providecommand{\urlprefix}{URL }
\providecommand{\doi}[1]{https://doi.org/#1}

\bibitem{aprosio2013extending}
Aprosio, A.P., Giuliano, C., Lavelli, A.: Extending the coverage of {DBpedia}
  properties using distant supervision over {Wikipedia}. In: NLP-DBPEDIA@ ISWC
  (2013)

\bibitem{bhagavatula2013methods}
Bhagavatula, C.S., et~al.: Methods for exploring and mining tables on
  {W}ikipedia. In: ACM SIGKDD Workshop on Interactive Data Exploration and
  Analytics. pp. 18--26. ACM (2013)

\bibitem{chu2012textual}
Chu-Carroll, J., Fan, J., et~al.: Textual resource acquisition and engineering.
  IBM Journal of Research and Development  \textbf{56}(3.4), ~4--1 (2012)

\bibitem{flati2014two}
Flati, T., et~al.: Two is bigger (and better) than one: the {W}ikipedia
  bitaxonomy project. In: 52nd Annual Meeting of the ACL. vol.~1, pp. 945--955
  (2014)

\bibitem{fleiss_kappa}
Fleiss, J.L.: Measuring nominal scale agreement among many raters.
  Psychological bulletin  \textbf{76}(5), ~378 (1971)

\bibitem{hearst1992automatic}
Hearst, M.A.: Automatic acquisition of hyponyms from large text corpora. In:
  14th Conference on Computational Linguistics. vol.~2, pp. 539--545. ACL
  (1992)

\bibitem{heist2018towards}
Heist, N.: Towards knowledge graph construction from entity co-occurrence. In:
  Doctoral Consortium at 21st International Conference on Knowledge Engineering
  and Knowledge Management (2018)

\bibitem{heist2017language}
Heist, N., Paulheim, H.: Language-agnostic relation extraction from {W}ikipedia
  abstracts. In: International Semantic Web Conference. pp. 383--399. Springer
  (2017)

\bibitem{heist2019uncovering}
Heist, N., Paulheim, H.: Uncovering the semantics of {W}ikipedia categories.
  In: International Semantic Web Conference. pp. 219--236. Springer (2019)

\bibitem{hertling2017webisalod}
Hertling, S., Paulheim, H.: Webisalod: {P}roviding hypernymy relations
  extracted from the {W}eb as {L}inked {O}pen {D}ata. In: International
  Semantic Web Conference. pp. 111--119. Springer (2017)

\bibitem{kalyanpur2012structured}
Kalyanpur, A., Boguraev, B.K., et~al.: Structured data and inference in
  {DeepQA}. IBM Journal of Research and Development  \textbf{56}(3.4),  10--1
  (2012)

\bibitem{kuhn2016type}
Kuhn, P., Mischkewitz, S., Ring, N., Windheuser, F.: Type inference on
  {W}ikipedia list pages. Informatik 2016  (2016)

\bibitem{lehmann2015dbpedia}
Lehmann, J., Isele, R., Jakob, M., et~al.: {DBpedia}--a large-scale,
  multilingual knowledge base extracted from {Wikipedia}. Semantic Web
  \textbf{6}(2),  167--195 (2015)

\bibitem{mahdisoltani2013yago3}
Mahdisoltani, F., Biega, J., Suchanek, F.M.: {YAGO3}: A knowledge base from
  multilingual {Wikipedias}. In: CIDR (2013)

\bibitem{mintz2009distant}
Mintz, M., et~al.: Distant supervision for relation extraction without labeled
  data. In: Joint Conference of the 47th Annual Meeting of the ACL and the 4th
  International Joint Conference on NLP of the AFNLP. vol.~2, pp. 1003--1011.
  ACL (2009)

\bibitem{munoz2014using}
Mu{\~n}oz, E., Hogan, A., Mileo, A.: Using linked data to mine {RDF} from
  {W}ikipedia's tables. In: 7th ACM International Conference on Web Search and
  Data Mining. pp. 533--542. ACM (2014)

\bibitem{paulheim2013extending}
Paulheim, H., Ponzetto, S.P.: Extending {DBpedia} with {Wikipedia} list pages.
  NLP-DBPEDIA@ ISWC  \textbf{13} (2013)

\bibitem{ponzetto2009large}
Ponzetto, S.P., Navigli, R.: Large-scale taxonomy mapping for restructuring and
  integrating wikipedia. In: 21st International Joint Conference on Artificial
  Intelligence (2009)

\bibitem{schrage2019extracting}
Schrage, F., et~al.: Extracting literal assertions for {DBpedia} from
  {Wikipedia} abstracts. In: International Conference on Sem. Systems. pp.
  288--294. Springer (2019)

\bibitem{topper2012dbpedia}
T{\"o}pper, G., et~al.: {DBpedia} ontology enrichment for inconsistency
  detection. In: 8th International Conference on Semantic Systems. pp. 33--40.
  ACM (2012)

\bibitem{wu2007autonomously}
Wu, F., Weld, D.S.: Autonomously semantifying wikipedia. In: 16th ACM
  Conference on Information and Knowledge Management. pp. 41--50. ACM (2007)

\end{thebibliography}
\end{document}